# Integrating AI for Enhanced Feedback in Translation Revision: A Mixed-Methods Investigation of Student Engagement

Simin Xu[1], Yanfang Su[1] & Kanglong Liu[1]*

[1]Department of Chinese and Bilingual Studies, The Hong Kong Polytechnic University, Hung Hom, Kowloon, Hong Kong SAR, China

**Author notes**

Correspondence should be addressed to Kanglong Liu, Department of Chinese and Bilingual Studies, The Hong Kong Polytechnic University, Hung Hom, Kowloon, Hong Kong SAR; Email: kl.liu@polyu.edu.hk

**Abstract**

Despite the well-established importance of feedback in education, the application of Artificial Intelligence (AI)-generated feedback, particularly from language models like ChatGPT, remains understudied in translation education. This study investigates the engagement of master's students in translation with ChatGPT-generated feedback during their revision process. A mixed-methods approach, combining a translation-and- revision experiment with quantitative and qualitative analyses, was employed to examine the feedback, translations pre- and post-revision, the revision process, and student reflections. The results reveal complex interrelations among cognitive, affective, and behavioural dimensions influencing students' engagement with AI feedback and their subsequent revisions. Specifically, the findings indicate that students invested considerable cognitive effort in the revision process, despite finding the feedback comprehensible. Additionally, they exhibited moderate affective satisfaction with the feedback model. Behaviourally, their actions were largely influenced by cognitive and affective factors, although some inconsistencies were observed. This research provides novel insights into the potential applications of AI-generated feedback in translation teachingand opens avenues for further investigation into the integration of AI tools in language teaching settings.

# 1. Introduction

Feedback is a crucial component of educational practice, with a profound impact on student achievement (Hattie, 1992). In the context of translation teaching, feedback plays a vital role in guiding students towards improving their translation skills (Bruton, 2007). However, the traditional model of feedback, where teachers manually craft responses to student work, is time-consuming and often places a significant strain on teachers (Shen et al., 2017; Guo et al., 2024). The emergence of artificial intelligence (AI) tools, such as ChatGPT, offers a promising solution to enhance teacher efficiency and provide high-quality feedback. Recent studies have explored the potential of ChatGPT to generate personalized, timely, and detailed feedback (Guo & Wang, 2023; Su & Lin, 2023; AlGhamdi, 2024; Banihashem et al., 2024). However, the majority of these studies have focused on writing, with only a few examining the application of ChatGPT in translation education (Cao & Zhong, 2023). Moreover, research has shown that the effectiveness of feedback is not solely determined by its quality, but also by how students engage with it (Handley et al., 2011). The dynamic interplay between feedback providers and students is a critical aspect of the feedback mechanism. However, there is a notable scarcity of research investigating learner engagement with ChatGPT translation feedback. This study seeks to bridge the existing gap by investigating how Master's students majoring in translation interact with ChatGPT translation feedback on cognitive, affective, and behavioural levels. By exploring the potential of ChatGPT in facilitating translation instruction, this research aims to contribute to the advancement of translation pedagogy and inform the effective integration of AI tools in educational settings.

# 2. Related Work
## 2.1 ChatGPT feedback in language learning and translation

Feedback, a widely employed instructional tool in classrooms, was initially conceptualized as information provided by teachers, peers, books, and texts in response to learners' work or performance (Hattie & Timperley, 2007). In the context of translation, feedback plays a crucial role in supporting and extending learning goals, enabling translators to view their work from the perspectives of readers or users, and fostering the skill of self-assessment (Washbourne, 2014). Previous studies have investigated the efficacy of various types of feedback in translation education. For instance, Yu et al. (2020) demonstrated the effectiveness of written corrective feedback in translation, revealing that students with low L2 proficiency tend to benefit from direct feedback. Similarly, Li and Ke (2022) found that peer feedback not only improves student performance but also enhances their capacity for evaluative judgments. While the benefits of translation feedback have been well-documented, the process of providing it is often time-consuming, and the increasing class sizes have exacerbated the workload of teachers (Banihashem et al., 2024; Er et al., 2021). Consequently, teachers face the challenge of balancing the quality and timeliness of feedback when assessing students' work (AlGhamdi, 2024). This dilemma has prompted researchers to explore automated approaches to providing translation feedback that can alleviate the burden on teachers while maintaining the quality of feedback (Han & Lu, 2023).

In this context, the emergence of ChatGPT has introduced a novel perspective on feedback provision. As a generative AI (GenAI) chatbot developed by OpenAI, ChatGPT was trained on a vast corpus of texts (Ekin, 2023). Its capabilities have been demonstrated in a range of tasks, including writing and translating (Herbold et al., 2023; Lee, 2023), making it a viable tool in foreign language learning. Several studies have explored the potential of ChatGPT in generating feedback by comparing the characteristics of feedback generated by ChatGPT and teachers. For instance, Steiss et al. (2023) conducted a comparative analysis of the quality of human and ChatGPT feedback on writing assignments, revealing that AI and human feedback exhibited distinct features. Guo and Wang (2023) found that when assessing students' writing, teachers primarily focused on generating content-related and language-related feedback, whereas ChatGPT feedback addressed three aspects (i.e., content, organization, and language) equally. This study also highlighted that teachers held both negative and positive perceptions towards this type of feedback. AlGhamdi (2024) employed a blinded approach to investigate how computing students responded to ChatGPT feedback after using both ChatGPT and human feedback in technical writing. The results showed that ChatGPT had the capacity to generate consistent and detailed feedback. While numerous studies have examined the use of ChatGPT feedback in writing contexts, research exploring the potential of ChatGPT to provide feedback for translation tasks remains relatively underrepresented. One notable exception is the study by Cao and Zhong (2023), which examines the effectiveness of feedback generated by ChatGPT and teachers by comparing students' revised translation drafts. However, the study's scope is limited to assessing the quality of the translations, leavingthe crucial aspects of students' perceptions of feedback and revision operations unexplored. The ways in which students engage with the revision process also remain unclear. Effective feedback is not solely determined by its content and quality, but also by how it is interpreted and internalized by learners (Nicol & McFarlane-Dick, p. 210). Therefore, to gain a comprehensive understanding of the potential of ChatGPT feedback, it is essential to investigate how students interact with and utilize it during the revision process.

## 2.2 Learner engagement with feedback

Learner engagement, a multifaceted construct encompassing emotional, cognitive, and behavioural dimensions (Fredricks et al., 2004), is a crucial factor in education. It is widely regarded as a key indicator of the extent to which students are committed to learning (Cheng et al., 2023). This concept is equally relevant to feedback, as its effectiveness is inextricably linked to student engagement (Winstone et al., 2017; Jørgensen, 2019).

To gain a deeper understanding of students' engagement with feedback, researchers have refined its analytical framework (Ellis, 2010; Han & Hyland, 2015; Zhang & Hyland, 2018; Zheng & Yu, 2018; Qian & Li, 2023). Specifically, cognitive engagement refers to the cognitive processes that learners employ in response to feedback (Ellis, 2010). This construct can be further categorized into three sub-components: awareness, cognitive operations and meta-cognitive operations (Han and Hyland, 2015). Awareness, which is the fundamental level of cognitive engagement, encompasses two key aspects: noticing and understanding. Noticing refers to learners' ability to discern the intention of feedback, while understanding demonstrates the degree to which learners can identify errors and provide accurate explanations. Previous research has measured cognitive operations by examining the macro strategies that learners use to respond to feedback (Pan et al., 2023), as well as the cognitive strategies employed to process feedback and generate revisions. Furthermore, meta-cognitive operations have been identified as actions that regulate mental effort, comprising two dimensions: monitoring and planning (Qian and Li, 2023). Specifically, monitoring is conceptualized as learners' ability to identify additional errors and inaccuracies beyond those highlighted in the feedback. Planning strategies, on the other hand, involve learners' prioritization when addressing feedback, which helps to reduce cognitive load. Although directly observing cognitive engagement is difficult, it can be measured indirectly through questionnaires and stimulated recall protocols (Chen, 2021; Philp & Duchesne, 2016). To overcome this challenge, the present study will adopt a multi-method approach, combining questionnaires, interviews, and revision records to capture learners' understanding of feedback, their application of cognitive operations, and their use of meta-cognitive operations during the revision process.

Affective engagement, also known as emotional engagement, encompasses learners' affective responses to feedback (Ellis, 2010). According to Han and Hyland (2015), this construct is characterized by the emotions experienced upon receiving feedback and revising one's work, as well as attitudinal responses towards feedback. Building on this concept, Zheng and Yu (2018) proposed a framework that distinguishes between three components of affective engagement: *affect* (learners' emotions and feelings), *judgment* (positive or negative evaluation of feedback), and *appreciation* (the perceived value of feedback). To measure affective engagement, researchers commonly employ questionnaires and self-report methods (Philp & Duchesne, 2016; Guo et al., 2023; Fan & Xu, 2020; Lee et al., 2023), which will be adopted in the present study.

Behavioural engagement is closely tied to the actions learners take in response to feedback (Zheng and Yu, 2018). This construct encompasses revision operations, which refer to the extent to which learners utilize feedback, as well as observable strategies employed to improve their work (Han & Hyland, 2015). For instance, Zhang (2017) used interview responses and revision time to illustrate how students engage with computer-generated feedback behaviourally when not pressed for time. Similarly, Tian and Zhou (2020) analysed textual changes between learners' initial drafts and revisions to indicate behavioural engagement. In line with these studies, the present research aims to measure behavioural engagement through three indicators: time spent on revisions, revision operations, and revision strategies.

Building on the aforementioned framework, a growing body of empirical research has explored how students engage with feedback, revealing complex patterns of engagement and diverse ways in which students interact with and respond to feedback. For instance, Zheng and Yu (2018) examined students' engagement with feedback in writing classes and found that engagement is closely tied to language proficiency, resulting in imbalances among the three dimensions of engagement. Similarly, Yu et al. (2019) investigated the engagement of master's students with peer feedback during second language writing, uncovering a complex relationship both within and across the three dimensions. More recently, Cheng and Zhang (2024) studied how students engaged with AWE (automated writing evaluation)-teacher feedback in writing tasks, finding that students exhibited deeper engagement both behaviourally and cognitively, while displaying similar levels of affective engagement compared to students who received only teacher feedback. Despite the growing body of research, the majority of studies have been conducted in the context of second language writing, with only a few scholars focusing on translation feedback. A notable exception is Zheng et al. (2020), which explored how students engaged with teacher translation feedback. This study highlighted the interplay among the three dimensions of engagement, revealing their mutual influence and the imbalances that exist among them. However, the study's small sample size (N = 3) limits its generalizability, underscoring the need for further research in this area.

## 3. The present study

As previously discussed, research on student engagement with feedback primarily concentrated on second language writing, while translation feedback has received limited attention. The recent advent of ChatGPT, an advanced technology with sophisticated natural language processing capabilities, offers a promising opportunity to provide feedback on translation assignments. ChatGPT's ability to generate fluent, detailed, and coherent feedback for student assignments in a short time (Dai et al., 2023, p.1) can assist teachers in assessing large classes and reducing their workload. However, the effectiveness of ChatGPT in translation teaching remains largely unexplored, and the nature of student engagement with its feedback in specific learning contexts is unclear.

Drawing on the framework of student engagement with feedback (Zheng and Yu, 2018), this study aims to investigate how students engage with ChatGPT-generated feedback on their translations during the revision process, encompassing cognitive, affective, and behavioural dimensions. Specifically, this study will address the following research questions:

RQ1: How do students engage with feedback provided by ChatGPT cognitively, affectively, and behaviourally?

RQ2: How do cognitive, affective, and behavioural engagement interact and influence each other when students respond to ChatGPT feedback?

## 4. Methodology
### 4.1 Participants

The study was conducted at a university in Hong Kong, with a sample of 29 students enrolled in the Master of Translating and Interpreting programme (MATI). The sample consisted of 21 female students (72.4%) and 8 male students (27.6%), which reflects the gender distribution in the MATI programme. Before the experiment, participants completed a pre-study survey to assess their familiarity with ChatGPT. The results showed that all students were aware of ChatGPT, and the majority (n = 23) had prior experience using it. Following a thorough explanation of the experimental procedure, all participants provided informed consent by signing a consent form.

**4.2 Data Collection**

Prior to data collection, our research team developed a specialized AI-powered Translation Teaching Platform, specifically designed to facilitate students' translation learning (see Fig. 1). The study utilized the platform's test function, which integrated ChatGPT to provide feedback. To ensure consistent and detailed feedback, a built-in prompt was created based on the translation assessment rubric from Hurtado Albir (2015) (see Fig. 2).

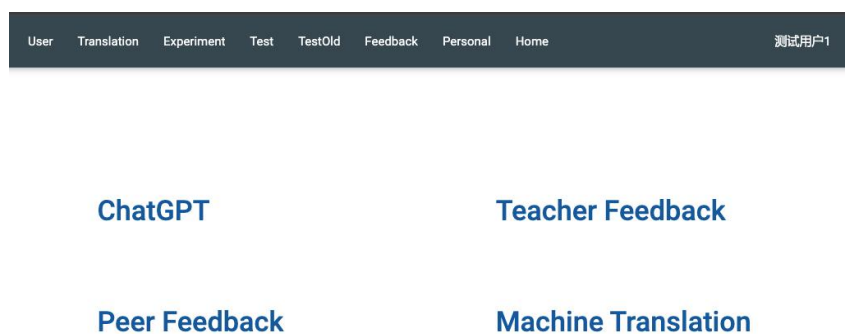

**Fig.1** The feedback function of the AI-powered Translation Teaching Platform

> Now you are a teacher in translation major. Here is a translation assessment rubric consisting of 3 aspects: Expression of the meaning of the original text accounts for 40% and it concludes same information, same clarity, and same register. Composition in the target language accounts for 40% and it concludes conventions of written language (correct orthography and typography), vocabulary (appropriateness and richness), morphosyntax (good use of verb tenses and modes, prepositions, etc.), cohesion (good use of connectors and referential elements), and coherence (ideas well organized and clearly presented). Level of communication of the target text accounts for 20% and it concludes overall quality of the target text, appropriateness of the genre's conventions, and appropriateness of the translation's purpose and target audience. Please give comments, suggestions and a score out of 100 (including the score of each aspect and the overall score) based on the above mentioned rubric for the following translated text: *translated text*. The source text is: *source text*.

**Fig.2** The built-in prompt provided to generate ChatGPT feedback

In the experiment, the participants were first required to translate a Chinese text of approximately 190 words within 70 minutes. After completing the translation, the platform generated feedback used ChatGPT on their performance. The participants then revised their translations based on this feedback. To match the teaching resources in the classroom, we selected a piece of political news as the source text. It is worth noting that, although ChatGPT can generate multiple versions of feedback based on the same prompt, we only considered the first outcome. Additionally, to ensure that all participants completed their initial translation draft without the aid of machine translation tools or online resources, they were required to share their screen on Zoom during the translation process. Videos were also recorded during the revision process to explore participants' behaviour.

Following the revision phase, all participants were required to finish a post-survey which focused on their engagement with ChatGPT feedback. The survey is designed based on Chen (2021)'s questionnaire which is originally designed to assess how students engaged with peer and teacher feedback. It consists of 15 Likert-scale questions with five response options: 1—strongly disagree, 2—disagree, 3—Neither agree nor disagree, 4—agree, and 5—strongly agree. All the questions in each dimension are divided into different sub-categories (See Table 3 and Table 4). To clarify the students' responses of the survey and amplify their revision processes in detail, we conducted follow-up interviews with a subset of participants after the experiment. Following purposeful sampling strategies (Patton, 2015, p. 406), we selected four participants for in-depth interviews. To maintain the anonymity of the students, we assigned them pseudonyms: Student 1, Student 2, Student 3, and Student 4. We chose them because their survey responses reflected the group's average, ensuring they could provide a representative illustration of the typical trends of all participants. Additionally, their voluntary consent to participate further legitimized their inclusion in the study. Drawing on the framework established by Zheng et al. (2020), the 13 interview questions (see Appendix 1) were designed to elicit (1) reflective commentary on the translation feedback and (2) retrospective accounts of the revision process. All interviews were conducted in Chinese to ensure linguistic and cultural authenticity. A research assistant collected and collated all relevant data, including students' translation drafts, revision drafts, feedback, post-survey responses, and revision recordings.

## 4.3 Data Analysis

The data analysis encompassed text analysis of participants' feedback and drafts, qualitative interviews, and quantitative surveys, complemented by observational analysis of revision recordings.This study employed a structured measurement framework, which assessed learner engagement across three dimensions: cognitive engagement, affective engagement, and behavioural engagement. To facilitate data analysis, we identified and segmented the data source according to the sub-categories of learner engagement outlined in Table 1, which was refined and adapted from existing engagement frameworks (Zhang & Hyland, 2018; Zheng & Yu, 2018; Qian & Li, 2023).

**Table 1.** Categories and sub-categories of student engagement with relevant sources

| Categories | Sub-categories | Relevant sources |
| --- | --- | --- |
| Cognitive engagement | Noticing and understanding feedback | Post-survey, interview data |
| | Meta-cognitive strategies | Post-survey, Revision recordings, interview data |
| | Cognitive operations | Post-survey, revision recordings, interview data |
| Affective engagement | Emotional responses | Post-survey, interview data |
| | Attitudinal responses | Post-survey, interview data |
| Behavioural engagement | Revision operations | Revision recordings, interview data |
| | Revision strategies | Revision recordings, interview data |
| | Revision duration | Revision time |

### 4.3.1 Analysis of student translations and ChatGPT feedback

We coded the content of ChatGPT feedback following the assessment rubric designed for the prompts, as well as the framework established by Tian and Zhou (2020). The feedback was categorized into two primary categories: surface-level and meaning-level feedback. Surface-level feedback referred to comments that did not involve changes to the underlying meaning, including corrections related to written conventions, genre conventions, word tenses, modes, and prepositions. Conversely, meaning-level feedback necessitates changes to the underlying meaning, including cohesion, coherence, translation accuracy, and lexical choices. To investigate how participants addressed feedback and revised their translations, we categorized their modifications according to whether they responded to the feedback, and calculated the feedback quantity, feedback uptake quantity, and feedback uptake rate. Furthermore, we observed and analyzed all participants' revision recordings to document the time spent on revisions, as well as the type and frequency of strategies employed during the revision process (see Table 2). This comprehensive analysis enabled a detailed examination of participants' revision behaviours.

**Table 2** Coding examples of one student' revision recording

| Student name | Revision time | Revision strategy | Times |
|---|---|---|---|
| Participant 1 | 0:51:21 | Online searching | 6 |
| | | Online searching | 7 |
| Participant 2 | 0:46:21 | Using corpora | 5 |
| | | Using dictionary | 2 |
| Participant 3 | 0:54:34 | Using dictionary | 6 |
| | | Online searching | 4 |

### 4.3.2 Analysis of post-survey and interview data

Regarding the post-survey, we assessed cognitive engagement through items 1-8 and affective engagement through items 9-15. Students' responses to the post-survey provided valuable insights into their experiences during the revision process and their perceptions of the feedback received. The oral interviews were transcribed verbatim using the automatic speech recognition app iFlyRec (https://www.iflyrec.com) and subsequently proofread manually by a research assistant to ensure accuracy. Two coders thoroughly read the transcripts multiple times to gain a comprehensive understanding of the content, and then conducted coding based on the framework outlined in Table 1. A qualitative analysis was employed to provide a nuanced and in-depth understanding of interviewees' cognitive, affective, and behavioural engagement. To ensure the trustworthiness of the coding, the two coders cross-referenced students' revision drafts and recordings to validate their responses and resolve any discrepancies. In cases of ambiguity or disagreement, discussions were held to reach a consensus and ensure inter-rater reliability.

## 5. Findings

Prior to presenting the findings on the three dimensions of engagement, it is essential to illustrate the structure and format of the ChatGPT feedback employed in the current study. Our analysis revealed that the feedback generated by ChatGPT, based on the prompt input, typically consisted of three primary components: (1) a reference translation, (2) comments and suggestions, and (3) a grade.

### 5.1 Cognitive engagement

To investigate how participants engaged with ChatGPT feedback cognitively, we operationalized cognitive engagement into three dimensions: (1) noticing and understanding the feedback, (2) meta-cognitive strategies to monitor the revision process, and (3) cognitive operations in mental activities to recall and utilize feedback. The post-survey items were categorized according to these dimensions (see Table 3). Overall, the mean scores of the eight survey questions ranged from 3.31 to 4.28, indicating that the participants exhibited a high level of cognitive engagement with ChatGPT feedback. With regard to understanding the feedback, the results of Q1 and Q2 suggested that most participants did not experience significant difficulties in comprehending and revising their work based on the feedback. However, there were also contrasting opinions. Four interviewees noted that while they could understand most suggestions, they found some meaning-level points confusing. For instance, ChatGPT feedback advised Student 1 to focus on coherence and cohesion, but failed to specify the exact errors, leaving her uncertain about how to improve the translation.

**Table 3** Descriptive statistics of the cognitive engagement

| Sub-categories | Items | M | SD |
| --- | --- | --- | --- |
| Noticing and understanding feedback | Q1: I could totally understand the feedback. | 4.00 | 0.93 |
| | Q2: After reading the feedback, I could easily make improvement. | 3.55 | 1.06 |
| Meta-cognitive strategies | Q3: When I revised my translation, I had to refer to the feedback repeatedly. | 4.07 | 0.88 |
| | Q4: I put much effort into revising my translation draft based on the feedback. | 4.28 | 0.96 |
| | Q5: I firstly checked the grade and then referred to the feedback. | 3.76 | 1.35 |
| | Q6: Compared with other issues, I tended to prioritize surface-level problems. | 4.10 | 0.82 |
| Cognitive operations | Q7: When I revised my translation, I totally followed the feedback. | 3.31 | 1.07 |
| | Q8: After receiving feedback, I first critically thought about the feedback and then revised my translation draft according to it. | 4.24 | 0.79 |

The mega-cognitive strategies employed by participants in both tasks primarily involved monitoring their mental effort, as well as practicing and planning the revision procedure. As indicated by Q3 and Q4, processing feedback required considerable effort from participants (M = 4.28), and they had to review the feedback repeatedly during the revision process (M = 4.07). This revealed an inconsistency between students' understanding of feedback and their mega-cognitive strategies. For instance, Student 3 reported that she spent much time on reading feedback because she wanted to comprehend it thoroughly.

Mental effort was also invested in monitoring translation accuracy, which was accompanied by corresponding revision actions. Four interviewees reported actively examining whether their revisions had improved the quality of their translations by reading through their work after completing the revision process. Another exampleis through self-correction. For instance, Student 1 made a revision that was not identified by ChatGPT:

*When I read the source text "布林肯与韩正会晤时毫不讳言,'我们有机会在最近两国高层接触的基础上前进,是一件好事'。"again during the revision process, I*

*realized that there are quotation marks in this sentence, which I didn't notice when I first translated it. Considering that the source text is news, I think using direct speech could make the translation more accurate and objective.*

Another meta-cognitive operation employed by participants is planning. According to Q5, students tended to prioritize checking their grades when reviewing feedback (M = 3.76). For instance, Student 3 reported that she first checked her grade upon receiving the feedback to gauge the overall quality of her translation. However, participants exhibited varying preferences when deciding on the priority of addressing feedback (SD = 1.35). In the interview, Student 2 revealed that she consulted reference translations while revising, and then chose to review the feedback comments, followed by checking the grade last. She expressed scepticism about the validity of machine-generated grades. The results of Q6 indicated that, during the revision process, students tended to prioritize correcting surface-level errors before addressing deeper, meaning-level issues (M = 4.10). Student 1 elaborated on her strategy:

*First, I went through the feedback and made corrections as I spotted issues like sentence structure and grammar. Once that was done, I tackled the deeper problems related to word choices and accuracy, which took a bit more effort.*

Regarding cognitive operations, judgment was employed to process ChatGPT feedback. The mean score for Q8 (M = 4.24) was higher than that for Q7 (M = 3.31). The results indicated that, although students expressed a moderate inclination to totally follow the feedback, they contradicted themselves by expressing doubts about the accuracy and correctness of the feedback during the revision process. This led to corresponding actions, such as making self-initiated changes and choosing not to adopt the feedback. This inconsistency suggests a conflict within their cognitive operations. Three interviewees mentioned that they tended to re-evaluate the accuracy of feedback points that differed from their own versions by consulting additional resources, such as Google and corpora.

**5.2 Affective engagement**

According to the conceptual framework in Table 1, affective engagement encompasses both emotional responses and attitudinal responses. The analysis of survey and interview data revealed a complex pattern of this dimension. In terms of emotional responses, students exhibited positive feelings (see Table 4), which is consistent with the findings on cognitive engagement. Specifically, the majority of students found the revision process enjoyable (M = 4.00), and ChatGPT translation feedback generally boosted their confidence and provided encouragement, motivating them to refine their translations further (M = 3.86). A key factor contributing to this positive affective response was the high grades generated by ChatGPT. Student 1 expressed surprise upon checking her grade, as she had not expected such a high mark when she submitted her initial translation. This unexpected outcome boosted her confidence to further revise the translation. However, two students expressed contrasting opinions. Student 2 reported that her feelings were not significantly influenced by the grade, explaining that teachers' scoring is based on comparing the work of all students and provides a more nuanced understanding of translation proficiency within the classroom context. In contrast, ChatGPT's scoring does not offer such comparative insight.

Furthermore, the descriptive statistics in Table 4 revealed that while positive feedback indeed enhanced students' confidence (M = 4.69), negative feedback did not have a significant impact on their motivations during revision (M = 2.93). All the interviewees acknowledged that they initially felt happy when receiving the feedback. However, they subsequently noticed that ChatGPT tended to generate an overabundance of positive feedback and expressed a preference for more critical suggestionsfor improving the translation. Student 1 remarked that upon first viewing the feedback, she felt quite confident because all the ChatGPT comments were positive. Nevertheless, she soon realized that they did not contribute to improving her translation skills.

**Table 4** Descriptive statistics of the Affective engagement

| Sub-categories | Items | M | SD |
| --- | --- | --- | --- |
| Emotional responses | Q9: I felt confident and encouraged in translation revision after reading the feedback. | 3.86 | 1.03 |
| | Q10: I enjoyed the revision process very much. | 4.00 | 0.89 |
| | Q11: Positive feedback makes me happy. | 4.69 | 0.47 |
| | Q12: Negative feedback makes me frustrated. | 2.93 | 1.46 |
| Attitudinal responses | Q13: I'm interested in the content of the ChatGPT feedback. | 3.97 | 0.98 |
| | Q14: I expect to receive the ChatGPT feedback. | 4.00 | 1.00 |
| | Q15: To what extent do you satisfied with the whole feedback? | 3.55 | 0.78 |

With respect to students' attitudes, the results showed a positive response to ChatGPT feedback. Overall, they thought the content of ChatGPT feedback was interesting (M = 3.97) and expressed a strong desire to continue utilizing it in future translation teaching (M = 4.00). The positive comments garnered from the interviews were largely attributed to students' expectations and interest in the format of this innovative feedback type, as well as the potential of ChatGPT in translation. For instance, Students 2 and 4 noted that ChatGPT was capable of producing more native-like expressions, while Student 3 reflected that the provision of reference translations accompanied by specific comments was more beneficial than receiving only one type of feedback.

Furthermore, students demonstrated a moderately favourable response towards the effectiveness of ChatGPT feedback (M = 3.55). Upon reviewing the responses from the four interviewees, it is noteworthy that they exhibited varying levels of satisfaction with different aspects of the feedback. Specifically, three students expressed particular satisfaction with the feedback on lexical choices, while two students were pleased with the suggestions regarding sentence structure. The comments of Students 2 and 3 aptly summarize the opinions of the majority:

*Student 2: ChatGPT gave me some suggestions for my sentence structure, and I think they could really improve the quality of my translation.*

*Student 3: I'm always worried that my translation was not native. When I looked over ChatGPT's feedback, I noticed that some of the suggested word choices fit much better with the style of news.*

However, they also expressed a certain degree of disappointment. Notably, two students acknowledged that some suggestions were unnecessary. Student 1 articulated her dissatisfaction with ChatGPT feedback:

*When it comes to the term "分歧", I initially used "division", but ChatGPT suggested changing it to "difference". I didn't think that was necessary.*

Another factor contributing to the students' disappointment was the general nature of the ChatGPT feedback. All interviewees reported that, although ChatGPT identified issues or provided specific corrections, it would be more beneficial if ChatGPT could generate more detailed guidance. This sentiment was echoed by Student 3, who stated in the interview:

*The feedback said my translation lacked clarity, but it didn't specify where or give me any suggestions on how to fix it.*

It can be inferred from her response that their disappointment can be partly attributed to the confusion they experienced while interpreting ChatGPT's feedback, suggesting a consistency between their cognitive and affective engagement.

### 5.3 Behavioural Engagement

Students' behavioural engagement can be examined through their revision operations, observable strategies used to enhance translation quality, and revision time. Feedback uptake and textual modifications indicate how students approached revising their work. As shown in Table 5, the majority of modifications were made in response to

ChatGPT translation feedback (756 items), with a smaller number of self-identified modifications (96 items). Student 1 recalled that, as ChatGPT did not identify all the problems in her translation, she began to identify them independently to improve the translation. It also reflected her dissatisfaction with ChatGPT feedback in the affective dimension. Table 6 illustrates the content of ChatGPT feedback and the four participants' revision operations in response to it. It is clear that ChatGPT primarily generated suggestions at the meaning level (477 items), achieving an uptake of 63.73%. Surface-level feedback followed, with fewer items (279) and a lower uptake of 47%. Despite finding surface-level feedback easier to process, students presented more confusion when interpreting meaning-level feedback; however, the uptake rate of meaning-level feedback was more than twice that of surface-level feedback, indicating a discrepancy between their cognitive and behavioural engagement. This discrepancy can be attributed to "the lack of feedback on genre-specific conventions and practical aspects of translation" (Student 3). Furthermore, when recalling the revision process in the interview, the most frequently mentioned details were related to the lexical level. Student 4, who positively commented on ChatGPT's proficiency and demonstrated deep affective engagement, recalled her revision process:

*I noticed that ChatGPT could effectively address lexical issues. For instance, when it came to the word "强调", I initially translated it as "underline". However, ChatGPT corrected it to "emphasize", which I hadn't considered during my translation. I thought this word fit better with the context of news.*

In contrast, Student 1, who reported low satisfaction with the feedback provided by ChatGPT and deemed some suggestions unnecessary, opted to retain certain original versions unchanged. The student's comments illustrate that affective engagement with ChatGPT feedback has an impact on the decision-making regarding uptake.

**Table 5** Summary of modifications in translations

| Revision type | Amount |
|---|---|
| Correction based feedback | 756 |
| Self-correction | 96 |
| Total | 852 |

**Table 6** ChatGPT feedback and learners' uptake rate

| Feedback content | Feedback quantity | Feedback uptake quantity | Feedback uptake rate |
|---|---|---|---|
| Surface-level | 279 | 130 | 47% |
| Meaning-level | 477 | 304 | 63.73% |
| Total | 756 | 434 | 57.41% |

The screen recordings revealed that students employed several observable revision strategies to enhance their translation quality. Four strategies were commonly used, as shown in Table 7. The most frequently employed strategy was online searching, which occurred 32 times in total. Three interviewees reported utilizing this strategy to address doubts about specific feedback points and correct self-detected problems simultaneously. For example, when Student 3 noticed that ChatGPT had corrected her initial phrase "pragmatic conversation" to "practical conversation," she critically evaluated whether it was a commonly used collocation by searching it on Google (Fig. 3). The second most frequently used strategy, employed 26 times, was proofreading, which aimed to ensure the accuracy and correctness of students' translations after modifications(Student 1, Student 2). According to the remarks of three interviewees, doubts about feedback also triggered students to consult corpora and dictionaries. Notably, Student 2, who expressed trust in ChatGPT feedback, did not consult any additional resources to verify the reliability of the feedback and appeared to copy it uncritically, suggesting that her behavioural and cognitive engagement remained relatively superficial.

**Table 7** Summary of revision strategies

| Strategy | Occurrence |
|---|---|
| Online searching | 32 |
| Proofreading | 26 |
| Using dictionaries | 18 |

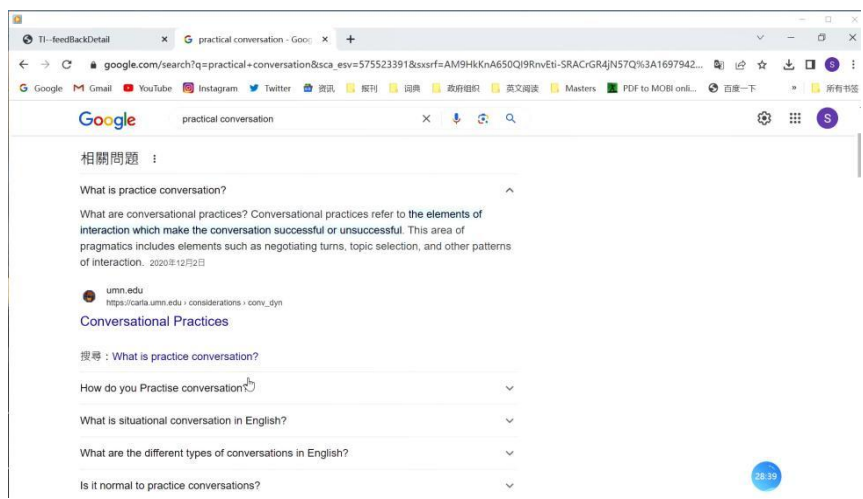

**Fig. 3** An example of Student 3' searching

Apart from the above-mentioned strategies, the interviews revealed that two participants would recall prior knowledge to make correction decisions and identify additional errors in the revision process. For example, Student 3 changed "extended his hope" to "in the expectation of", which was not included in ChatGPT's feedback. She explained her reasoning for this correction:

*When it comes to the word "希望", I initially translated it into "extend his hope". But as I was revising, I remembered something from my interpreting classes: "extend" is usually paired with "gratitude", not "hope". So I ended up changing it to "in the expectation of".*

The revision operations and strategies reveals a strong link between students' cognitive and behavioural engagement. Cognitive engagement appears to drive behavioural engagement, initiating and guiding students' actions during the revision process (Fan & Xu, 2020). This implies that students' thinking and critical evaluation directly impact their revision behaviours, highlighting the interaction between mental processes and actions in translation revision.

We categorized students' revision duration, which refers to the time spent on searching and textual modifications, into four ranges: less than 5 minutes, 5-20 minutes, 20-40 minutes, and more than 40 minutes. Fig. 4 illustrates the distribution of time per category. Notably, the majority of students spent between 5-20 minutes (41% of participants) and 20-40 minutes (31% of participants) on revision. Table 8 shows that students took an average of 24:41 to complete their revisions. It is worth highlighting that while the minimum revision time recorded was just 01:30, which is exceptionally brief, all other students' revisions lasted longer than 08:30. The unusually short revision time of this student might be attributed to a lack of motivation or interest in the task.

Three interviewees reported spending considerable time executing search queries on the internet. Interestingly, Student 3, despite expressing satisfaction with ChatGPT's lexical choices, spent 54:34 on revision—the maximum time among all participants. Much of this effort was devoted to verifying the correctness of ChatGPT feedback It reveals a discrepancy between her affective engagement and behaviours:

*Whenever I noticed that ChatGPT's suggestions differed from my translation, I would hesitate about whether to follow them by doing some searching using Google. Besides that, I would also proofread my revised translation to catch any other potential errors.*

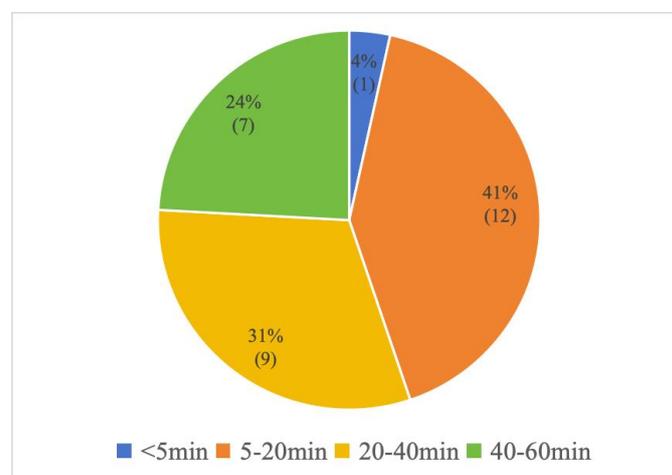

**Fig. 4** Time spent on revision

In contrast, Student 2, who also expressed trust in ChatGPT's feedback, accepted most of the content superficially without consulting any additional resources. This resulted in a relatively brief revision time of 09:45. The contrasting behaviours and motivations for conducting searches observed in these two students demonstrate individual differences in the mutual influence of various types of engagement.

**Table 8** Summary of students' revision time

| Mean | min | max | SD |
|---|---|---|---|
| 0:24:41 | 0:01:30 | 0:54:34 | 0:15:17 |

## 6. Discussion

This study revealed that students generally engaged actively with ChatGPT translation feedback on cognitive, affective, and behavioural levels, which supports the idea that collaboration with AI tools can positively impact language learning (Sahari et al., 2023). However, students' engagement demonstrated considerable complexity, illustrating intricate influences and connections both within sub-constructs of each dimension and across the three different dimensions.

Regarding cognitive engagement, participants generally found the feedback easy to understand, particularly surface-level suggestions. However, a small number of students struggled to comprehend meaning-level feedback, especially concerning cohesion and coherence, and required additional guidance. This difficulty arose from the overly general nature of the feedback, which hindered students' ability to process it effectively. This finding aligns with Su et al.'s (2023) research, which noted that ChatGPT sometimes tends to generate vague feedback. Additionally, students employed a variety of meta-cognitive strategies to regulate their mental effort and plan the revision procedure when addressing the feedback and making corrections. It indicated that students invested extensive cognitive effort to ensure the appropriateness of the feedback and the accuracy of their revised translations. Specific strategies and behaviours were activated, including repeatedly checking feedback, proofreading revisions, and self-correction. The considerable effort required to interpret ChatGPT's feedback aligns with the research of Zheng and Yu (2018), which suggests that processing general feedback demands greater linguistic competence from learners for accurate interpretation. The mismatch between feedback comprehension and meta-cognitive strategies highlights the inconsistency in sub-constructs of cognitive engagement. Students exhibited diverse preferences when deciding which parts to prioritize upon reviewing the feedback. As observed in the study, most students showed a tendency to first check their grades and assess whether their performance was satisfactory. This behaviour can be attributed to their competence preference associated with learning in traditional classroom contexts (Huguet et al., 2001; Elliot et al., 2018; Cassidy, 2008; Riemer & Schrader, 2022). Regarding the detailed feedback content, students addressed surface-level issues first. Compared to resolving meaning-level issues, these required lower cognitive investment (Yu et al., 2019) and less linguistic competence (Chandler, 2003). Furthermore, conflict was also observed within cognitive operations, aligning with the research of Jiang and Yu (2022), who noted that students had certain reservations when using automated feedback. They engaged in behaviours such as consulting additional resources to determine whether to incorporate the feedback into their work. This finding confirms ChatGPT feedback's ability to prompt students to proactively address their doubts, thereby enhancing their critical thinking skills—a crucial capacity for translators.

Students' affective engagement with ChatGPT feedback presented a complex picture, consistent with their cognitive engagement. They expressed enjoyment in the revision process and confidence upon checking the high grades and positive comments generated by ChatGPT. However, while students initially chose to check their grades upon receiving the feedback, some placed more emphasis on the feedback content . This preference for feedback over grades may be attributed to the fact that ChatGPT's scoring did not compare individual translation performance against that of the entire class. This observation suggests students' natural tendency to compare their grades with peers in learning activities, mirroring the findings related to cognitive engagement. Notably, students seemed to prefer critical feedback that provided more guidance for improvement, enhancing their motivation to complete revisions. This finding contradicts the research of Ilies et al. (2007), which suggested that negative feedback could damage students' confidence and affect their mood. Overall, students held generally positive views towards ChatGPT in the process of knowledge acquisition (Sallam et al., 2023). They valued the unique feedback structure as well as ChatGPT's capacity to address certain issues, especially in lexical choices and sentence structure. However, some students expressed frustration regarding the effectiveness of ChatGPT feedback, as not all suggestions were deemed necessary. This indicates that the confusion in comprehending ChatGPT feedback in the cognitive dimension can influence students' attitudes towards it.

Behaviorally, the revisions deploying ChatGPT feedback were not very successful. It can be attributed to previous research suggesting that learners are selective in their adoption of automated feedback, adjusting their uptake accordingly (Bai & Hu, 2017; Qian & Li, 2023; Storch & Wigglesworth, 2010). Notably, students tended to identify issues not detected by ChatGPT and made self-initiated changes, which may stem from their inherent scepticism about the accuracy of automated feedback, as reported in Fan and Xu (2020). This phenomenon highlights the complex relationship between cognitive engagement and behavioural engagement. Regarding specific aspects of feedback, despite students' reported difficulties with the understanding of meaning-level feedback and their tendency to prioritize surface-level issues, the majority of modifications were made in response to meaning-level feedback, with a higher uptake rate. This discrepancy between behavioural and cognitive engagement is consistent with Zhang et al.'s (2023) findings. A key influencing factor may be that while ChatGPT excels at providing contextually suitable vocabulary, it often fails to generate sufficient and explicit suggestions regarding genre conventions, thereby limiting students' ability to effectively utilize surface-level feedback. This limitation supports Zheng and Yu's (2018) claim that the feedback provider's practice can significantly influence students' revision behaviours.

Furthermore, our interview results revealed that the incorporation of feedback and decision-making processes were influenced by students' affective engagement

(Yu et al., 2019). However, the analysis of revision periods and detailed revision actions suggested that the interplay among the three dimensions of engagement is complex and varies due to individual factors, such as prior knowledge, learning styles, and motivational orientations (Afifi et al., 2023). For instance, students with higher levels of prior knowledge in the subject matter may be more likely to engage cognitively with the feedback, while those with lower levels of prior knowledge may rely more heavily on affective engagement, such as their emotional responses to the feedback. A case in point is Student 3, who expressed satisfaction with the feedback but still had doubts about its reliability. Notably, she invested considerable time in comprehending the feedback, demonstrating deep cognitive engagement, which in turn led to substantial behavioural engagement. Additionally, students devoted effort to verifying the reliability of ChatGPT feedback and subsequently employed strategies to address the feedback and regulate translation accuracy. This finding aligns with Wang et al.'s (2022) study, which highlighted the impact of cognitive processing on behavioural engagement. The complex interplay among the three dimensions of engagement underscores the need for a nuanced understanding of how students interact with automated feedback, taking into account individual differences and contextual factors.

Several pedagogical implications can be drawn from this study. First, feedback itself cannot lead to learning gains unless students generate internal feedback through their own processing of the information provided (Yan & Carless, 2022). Students in this study reflected that the general nature of the feedback prevented them from processing it effectively, underscoring the importance considering how the characteristics of feedback impact student engagement and influence their revision processes. Second, it is notable that students prefer to utilize more critical suggestions over positive feedback. Accordingly, teachers should adopt a holistic approach when designing and adjusting feedback strategies based on individual engagement and performance (Shen & Chong, 2023). Specifically, the design of feedback should ensure a balanced inclusion of both recognition of strengths and constructive critical comments. The study found that students tended to employ cognitive operations to decide whether to accept feedback and utilized meta-cognitive strategies to address feedback and make revisions. However, managing the quality of translation and processing feedback information can be challenging for students, particularly when their language proficiency and feedback literacy are limited (Zhang, 2020). To address these challenges, additional instruction should be incorporated into daily teaching, including guidance on developing linguistic skills, strategies for applying feedback effectively, and techniques for error detection (Zhang, 2017; Malecka et al., 2022). Finally, students' engagement patterns suggest that ChatGPT performs satisfactorily on basic levels, including lexical choices and sentence structures, while showing moderate performance on higher-level issues, such as cohesion and coherence. In future translation teaching, ChatGPT feedback can be effectively integrated with teacher feedback to handle basic-level concerns. This approach allows teachers to concentrate on providing more nuanced feedback, ensuring comprehensive coverage and potentially reducing their workload, especially when dealing with the practical challenges of large class sizes (Guo et al., 2024). However, to successfully integrate AI into classroom practices, it must be developed with a clear understanding of educational regulations (Kim, 2023). Educators should also enhance their expertise in prompt engineering to improve the efficacy of ChatGPT outcomes.

## 7. Conclusion

The findings of this study underscore the transformative potential of ChatGPT-generated feedback in the realm of translation education. By introducing AI into the feedback loop, educators can not only alleviate the heavy workload traditionally associated with manual feedback but also enhance the quality and timeliness of the feedback provided to students. This integration can lead to a more dynamic and responsive learning environment where students receive immediate and detailed insights into their translation work. Moreover, the study highlights the importance of fostering a robust engagement framework that encompasses cognitive, affective, and behavioural dimensions. A nuanced understanding of how students interact with AI-generated feedback can inform the development of more effective pedagogical strategies. For instance, tailoring feedback to address individual learning needs and preferences can significantly enhance student satisfaction and learning outcomes. Furthermore, the implications of this research extend beyond translation education, offering valuable insights for the broader field of language learning and teaching. Future studies should explore the scalability of ChatGPT feedback in different educational contexts and disciplines, as well as its long-term impact on student learning trajectories.

Despite these findings and the valuable insights gained, our study had some limitations that should be acknowledged. First, since the output of ChatGPT is influenced by the quality of prompts (OpenAI, 2022), the effectiveness of feedback could be improved by enhancing prompt quality. Second, we utilized ChatGPT 3.5, the most advanced version available at the time of the experiment. As technology progresses, more recent AI models should be considered to ensure the study remains current. Third, our investigation of student engagement was limited to a single experiment, which may not fully capture the complexities of engagement over time. Future studies should explore the long-term impact of ChatGPT feedback on student learning by conducting longitudinal research. Additionally, comparative studies involving different AI models could provide a deeper understanding of the evolving capabilities of AI in educational contexts. Further research should also investigate the potential of ChatGPT feedback across diverse learner populations, including those with varying levels of proficiency and training experience, to determine its broader applicability in translation education.

**Data Availability**

The data that support the findings of this study are available from the corresponding author upon reasonable request.

**Declarations**

**Ethical Approval**

The authors confirm that all the research meets ethical guidelines and adheres to the legal requirements of the study country.

**Consent to Participate**

Informed consent was obtained from all participants involved in this study.

**Conflict of Interests**

The authors declare no conflict of interests.

**Appendix 1. Guide for the Semi-Structured Interview**

1. What actions did you take upon receiving feedback? Could you describe them in detail?
2. How did you feel after receiving your first feedback?
3. How did you feel after receiving your second feedback?
4. Did you feel encouraged when reading the feedback?
5. Please compare the feedback on the two drafts and tell me which one you think is better, and why?
6. What are your thoughts on the holistic scores and feedback?
7. Do you find the feedback helpful?
8. Is the feedback easy to understand?
9. Did you encounter any problems with the feedback?
10. When you encountered problems with the teacher's feedback, what did you do?
11. Can you identify the strengths and weaknesses of ChatGPT feedback?
12. Do you think ChatGPT feedback should be used in translation education? Why?
13. How do you think ChatGPT feedback could be utilized in translation teaching?